\documentclass[runningheads]{svmult}

\usepackage{makeidx}   
\usepackage{graphicx}  
\usepackage{subeqnar}  
\usepackage{multicol}  
\usepackage{physprbb}  
\makeindex             

\usepackage{amsmath}   
\begin{document}

\title*{The art and science of black hole mergers}
\toctitle{The art and science of black hole mergers}
\titlerunning{Black hole mergers}

\author{Bernard F Schutz\inst{1,}\inst{2}}
\authorrunning{Bernard F Schutz}

\institute{Max Planck Institute for Gravitational Physics  
        (Albert Einstein Institute),
        14476 Potsdam Germany 
\and Department of Physics and Astronomy, 
     Cardiff University, Wales}
\maketitle              

\begin{abstract}
The merger of two black holes is one of the most extraordinary events in the natural world. Made of pure gravity, the holes combine to form a single hole, emitting a strong burst of gravitational radiation. Ground-based detectors are currently searching for such bursts from holes formed in the evolution of binary stars, and indeed the very first gravitational wave event detected may well be a black-hole merger. The space-based LISA detector is being designed to search for such bursts from merging massive black holes in the centers of galaxies, events that would emit many thousands of solar masses of pure gravitational wave energy over a period of only a few minutes. To assist gravitational wave astronomers in their searches, and to be in a position to understand the details of what they see, numerical relativists are performing supercomputer simulations of these events. I review here the state of the art of these simulations, what we have learned from them so far, and what challenges remain before we have a full prediction of the waveforms to be expected from these events.
\end{abstract}

\section{Black-hole coalescence systems}
Black holes are the ultimate in strong gravity, and the details of their merging require general relativity for any kind of even approximate description. Nevertheless, it is one of the remarkable consequences of general relativity that, during the orbital phase before coalescence, the black holes follow orbits that are described to first order by Newtonian gravity: their interaction when separated by a significant distance does not reflect the enormously strong gravity inside and near them. Only when they come within a few tens of gravitational radii do we require full general relativity to describe the dynamics.

Before that, the post-Newtonian approximation -- an asymptotic approximation to general relativity valid for small orbital velocity ($v/c \ll 1$) in gravitationally bound systems -- provides a systematic approach to studying the orbital inspiral phase, where orbits shrink and lose eccentricity through the radiation of energy and angular momentum in gravitational waves.\cite{BlanchetLR} The classic test of gravitational wave theory, the Hulse-Taylor binary neutron-star system, is very accurately described by such an approximation.\cite{StairsLR} 

The gravitational radiation emitted by orbiting black holes comes out at multiples of the orbital frequency $f_\text{orb}$, starting at $f_\text{gw}=2f_\text{orb}$. Higher harmonics are important only if the orbit is highly eccentric. Other frequencies, including $f_\text{orb}$ itself, can appear in the spectrum from the coupling of black-hole spins to the orbital angular momentum and to each other, but this is significant only when the holes are very close to one another. When they are well separated, the orbital frequency is the Newtonian one, 
\[f_\text{orb} = \frac{1}{2\pi}\left(\frac{GM}{R^3}\right),\]
where $M$ is the total mass and $R$ the separation of the holes. Putting in some numbers, this gives for the dominant gravitational wave frequency
\[f_\text{gw}=2\times10^{3}\text{Hz}\left(\frac{M}{10M_\odot}\right)^{-1}\left(\frac{R_g}{R}\right)^{3/2},\]
where $R_g$ is the gravitational radius (radius of the horizon) of a black hole of the total mass $M$, {\em i.e.}\ of the final merged black hole.  

From this it is clear that ground-based detectors, which are most sensitive around 100~Hz and which will eventually reach down to around 10~Hz, will be looking at mergers of black holes produced by stellar evolution, up to about $100M_\odot$. Even black holes formed in the first generation of stars, which could have masses of a few hundred solar, may be visible to advanced detectors only in the final merger event, but not during the long inspiral phase. Most merger events in the ground-based frequency window probably result from the long-term evolution of stellar binary systems, in which the mass-ratio between the holes is probably not very large and the orbits have circularized early. 

For the LISA gravitational wave detector, the most sensitive frequency window is between 1 and 10~mHz, which places the masses of the holes in the $10^4$--$10^6M_\odot$ range: black holes in the centers of galaxies. For these, two kinds of merger events that seem promising sources and have received the most attention from theorists.\cite{CutlerThorne} In one scenario, two comparable-mass black holes are brought together by dynamical friction with the background stellar population. Here the orbital eccentricity during the observation period might be very small. 

In the second scenario, a compact object like a neutron star or (even better) a $10M_\odot$ black hole is placed by random collisions into a plunge orbit that takes it close enough to the black hole to be captured on the first pass, due to the loss of energy to gravitational radiation. After that it may take millions of orbits before its distance of closest approach inches close enough to the horizon for it finally to get captured. Studies suggest that in this case the orbit remains substantially eccentric right up to the end. 

These two scenarios require very different techniques to study them theoretically. In the capture scenario, the mass ratio should be smaller than 0.1\%, which means that the inspiralling hole can be treated as a small perturbation on the background of the larger hole. This perturbation study is a hot topic in theoretical research today and is far from solved. However, the problem has been understood well enough for theorists to have confidence that LISA will see hundreds of these events.\cite{Gair} Given that the inspiralling object spends so much time probing the geometry near the central black hole, these capture events will provide an ideal test of the Kerr metric and general relativity's theorems about black-hole uniqueness.\cite{CutlerThorne} 

The first scenario, where two holes merge with a mass ratio that is not very different from unity, must be studied by numerical techniques. Post-Newtonian orbital approximations may apply up to very small separations ($R/R_g \sim 10$?), but the final plunge and merger phase must be studied numerically. This is another hot topic in relativity theory. The focus of this article is to describe what we have learned about this process so far.

\section{The three phases of black hole merger}
Provided that the black holes start far enough apart, the lifetime of the system is dominated by the lowest-order radiation reaction corrections. From this we learn that the time to coalescence of a circular binary whose component masses are $m_1$ and $m_2$, whose total mass is $M=m_1+m_2$, and whose initial separation in a circular orbit is $R$, is
\begin{equation}\label{eq:lifetime}
t_c = \frac{5}{256\delta(1-\delta)}\frac{R}{c}\left(\frac{GM}{c^2R}\right)^{-3}, 
\end{equation}
where $\delta$ is the {\em mass fraction}:
\[ \delta = m_1/M.\]
Solving for the separation, we see that for holes to coalesce due to gravitational radiation reaction within a time short compared to a Hubble time, their separation must be small:
\begin{equation}\label{eq:critsep}
R_c = 2\times10^{-3} \delta^{1/4}(1-\delta)^{1/4}\left(\frac{t_c}{10^{10}\text{yr}}\right)^{1/4}\left(\frac{M}{10^6M_\odot}\right)^{3/4}\quad\text{pc}.
\end{equation}
The stellar dynamics in the central star cluster that are required to bring holes this close are discussed elsewhere in this volume by a number of authors.

\begin{figure}[b]
\begin{center}
\includegraphics[width=.6\textwidth]{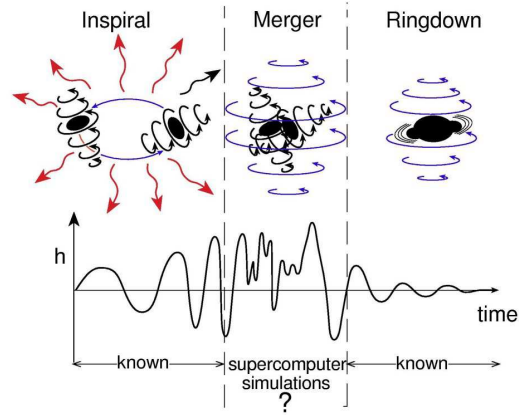}
\end{center}
\caption[]{The three phases of black hole merger (courtesy Kip Thorne).}
\label{fig:threephases}
\end{figure}
The black holes spend almost all their time, and (for $10^4$--$10^6M_\odot$ holes) almost all the time they are visible to LISA, in this first phase of the dynamics, where the stars orbit one another and gradually spiral together. 

Eventually the post-Newtonian description of the orbit breaks down, and the holes cannot be treated as point particles any more. What is more, it is expected that the holes will reach the {\em innermost stable circular orbit} (ISCO), at which the gradual inspiral ends and the holes plunge together. This is what happens to point particles in the Kerr geometry: the location of the ISCO depends on the spin of the hole, and it can be anywhere from 3 times the horizon size (non-spinning Schwarzschild hole) to just above the horizon (maximally spinning Kerr hole). Once the holes begin to plunge together, they rapidly meet and merge into a single black hole. This second phase of the merger process cannot be understood analytically and is the object of the numerical work that I will describe below. We don't even have a very good idea of where the transition occurs between the first and second stages.

The final stage of the merger is better understood: very quickly the final black hole radiates away all its non-axisymmetry and settles into a Kerr black hole. The radiation is dominated by the excited quasi-normal modes of the black hole, which have strong damping: within a few cycles (a few light-crossing times) the hole is substantially axisymmetric. This process is understood at the level of perturbation theory of black holes and has been studied numerically, with no surprises.

\section{Why study mergers numerically?}
Of these phases, the one that is least understood is the second. At the moment, numerical studies only handle black holes that are very close to one another, near the ISCO. Studies have been able to follow holes from such an orbit through to merger and ringdown, but we do not yet know if the starting conditions are physically appropriate. I will come back to this key issue below. The computer resources needed to study this problem are substantial. Why, then, should we be trying so hard to understand this process?

There are at least four reasons:
\begin{enumerate}
\item Gravitation theory. Although we have good theoretical reasons for believing that general relativity's description of time-independent black holes is correct, we do not know what the theory says in detail about dynamical horizons and mergers. The process is strongly nonlinear, and it appears therefore that numerical experiments are the only way of doing {\em experiments} in this regime.
\item Simplest dynamical problem in general relativity. Black hole mergers are in some sense the least complicated strong-field process, since they are not complicated by the need to include fluid dynamics, atomic or nuclear physics, or magnetic fields. They do have a complication that stars do not: the singularity at the center. This challenges the numerical formulation of the problem, but it is nevertheless true that the ``parameter space'' of merger models will be small compared to that of stars. 
\item Gravitational wave detection. One of the principal motivations for studying this problem is the need for good numerical waveform predictions as an aid to the detection of these events. Ground-based detectors will be operating in one year at the level of sensitivity where the first black-hole coalescences could potentially be seen, but only if the signal waveform can be predicted well enough. Waveform predictions are used to construct {\em matched filters} for the data, which find weak signals buried in the noise if the incoming waveform correlates well with the predicted one. For the first stage of inspiral, our predicted waveforms are adequate. But much of the signal is expected to come from the second stage, and so there is considerable pressure from the detector community to make useful predictions. And when signals are detected, the predictions will be needed for their interpretation: what do they say about the mass ratio and spins of the holes? Finally, for LISA, there is an additional problem, that the signals will be so strong that they will contaminate the data and must be removed in order to search for weaker signals, such as those from the capture of smaller black holes. This removal can only be done adequately if there is a good waveform prediction.
\item Astrophysically interesting predictions. Simulations could help fill in the answers to a number of important astrophysical questions. The angular momentum radiated in the second phase (plunge and merger) determines the spin of the final black hole. Can holes with extreme spin be formed by mergers? The linear momentum radiated determines the recoil velocity of the final black hole. Will such holes remain bound in their star clusters or galaxies after merger? (See the article by Scott Hughes in this volume.)
\end{enumerate}

\section{Practical numerical simulations: what are the issues?}
A glance at some numbers will show why the simulation of the merger of two black holes is a challenging task. 

Grid meshes must be large but at the same time fine. The horizon must be well resolved, requiring a fine mesh. But the speed of the black holes in their orbits before the plunge is actually nonrelativistic, so the gravitational wavelength is long. To resolve the waves, one needs to place the grid boundary several wavelengths away. In three dimensions, this can be costly. Using the mass $M$ as a distance unit (in ordinary units this is $GM/c^2$), a mesh resolution of $0.01M$ and an outer boundary distance of $100M$ (which is only $1.5\lambda$ if the initial orbit is the Schwarzschild ISCO) already requires $10^{12}$ grid points.

In addition, there is a big memory requirement per grid point, because modern formulations of the field equations, specially designed for numerical stability, require between 50 and 80 double-precision variables per grid point. A single time-step could therefore require 400 to 600 TB of main memory to store the mesh described above. 

The nonlinearity of the Einstein field equations means that computing the functions needed to advance the simulation by one time-step is costly. The equation for each variable can involve all the others at neighboring points (7 points in second-order-accurate methods, many more in more accurate schemes), in non-linear combinations (polynomials, matrix inversions). The result can require thousands of floating-point operations per variable per grid point per time step. A single time-step on the mesh described above could demand $10^{17}$ operations. A teraflop computer would take a day to do just one time-step!

Finally, there need to be a lot of time-steps. The time-step size is set by the Courant condition, at least in the explicit finite-difference methods used by the AEI group and most others. Since the horizon must be resolved with hundreds of cells, a single orbit will traverse thousands of grid cells, and will therefore require thousands of time-steps. A single orbit on our hypothetical mesh and hypothetical teraflop computer would therefore take three years to compute!

It is clear from these numbers that calculations done at present must make compromises or use clever tricks to get around these problems. I will discuss these approaches in the next section. But first it is important also to understand that the problems facing groups doing these simulations are not just brute-force computing power issues. They also involve a host of subtle, theoretical issues associated with Einstein's equations.

For example, inside the black holes there are singularities. These must be avoided somehow, since the computer cannot compute accurately near them. The method of choice today (a big advance over even 5 years ago) is excision: a region of the grid mesh containing the singularity is simply left out of the numerical integration. This is possible because, as long as this region is entirely contained inside the horizon, no errors made in the solution inside the horizon can leak out and affect the solution outside: the interior of the black hole is causally decoupled from the exterior. However, that is only true of the exact Einstein equations. Their finite-difference approximation may well be able to transfer information (and errors) from inside the horizon to outside. So excision needs to be handled with care, and at present there is a strong suspicion that some slowly growing instabilities seen in codes that use excision may be caused by non-causal effects in the finite-difference equations. 

Most formulations of Einstein's equations restrict the integration to a finite domain, so not only is there an inner boundary at the excision surface, but there is also an outer boundary at the edge of the grid. Remarkably, it is not known how to set an accurate boundary condition there. Unlike the simple case of a scalar wave equation, where an outgoing wave boundary condition is simple to impose and does a good job of imposing causality on the system, in general relativity there are too many wave-like degrees of freedom and many non-radiative variables, all coupled together. So far, the only remedy has been to try to put the boundary and its over-simplified boundary condition so far away that unwanted reflected waves and information arrive back in the center of the integration domain only after the most interesting events (plunges, mergers) have happened.

What is more, there is not even a unique formulation of the Einstein equations. Coordinate freedom means that one can change the variables one integrates and the way they couple to one another, and one can even change the slicing of spacetime into space sections. And among Einstein's equations there are four so-called constraint equations that have an elliptical structure. This means that it is possible to insert them into the remaining six dynamical equations in any number of ways (via linear combinations or more complicated insertions) without changing the physical solution of the equations. This freedom has been seen to have a big effect on the stability of the integration methods, and therefore is a useful but complicating measure.\cite{BSSN}

Once one has some output from a numerical simulation, there remains the important challenge of interpreting it: finding horizons, discovering causes of instabilities, visualizing the physical and geometrical variables. The AEI group has put a great deal of effort into this problem, producing along the way many beautiful images and movies. But it is fair to say that the subject of visualization of three-dimensional simulations in general relativity is in its infancy and could reqard a systematic study.

Perhaps the most troubling problem of all, at least at present, is the correct formulation of the initial data for a numerical integration. I will describe this in detail in a later section.

All of this makes clear that the problem is a complex one, and there is a premium on collaboration. Many different approaches need to be explored, but at the same time one needs a framework in which they can be compared. Moreover, there is much commonality in the computational infrastructure required for many of these otherwise different approaches. 

The AEI group recognized this long ago and began to develop the Cactus Computational Toolkit to enable teams of scientists to collaborate more effectively, even over large distances.\cite{Cactus} Recently a group of leading scientists working in numerical relativity, most of them Cactus users, formed a quality-assurance organization called Apples with Apples.\cite{apples} The aim is develop a set of specific tests that could be applied to all variants, all approaches to the numerical relativity problem. This then allows one to compare ``apples with apples'' rather than ``apples with pears'' in deciding which approach is more promising. 

\section{Addressing the issues}
I described the computer-power challenges in the previous section as starkly as I could, but it is clear that there are methods to reduce the size of the problem. All the demands are driven by the size of the grid mesh. This can be reduced in a number of ways. Mesh refinement is an obvious one, in which there is not a uniform mesh spacing everywhere. Near the horizon one needs finely spaced grid points, but these are wasted far away. Groups have now begun to do numerical integrations in meshes consisting of a series of concentric boxes, with coarser and coarser resolution as one goes out.\cite{Hawley,Centrella} This can dramatically reduce the number of grid points by factors of 1000 or more. Then a three-year integration becomes the work of one day. 

Even more radical solutions are possible in the framework of spectral methods, which dispose of grid points in favor of representing variables by sums of basis functions, whose coefficients become the numerical variables of the problem. This global approach offers considerable promise, not only in reducing the size of the problem but also in curing instabilities.\cite{Cornell,Meudon}

Studies of singularity excision are beginning to understand how they might create weak instabilities. Excision seems to work best, at least at present, in comoving coordinates, where the black holes do not move across the mesh but rather the mesh follows them as they merge together. The coordinate freedom of general relativity permits this, and in fact the fixed mesh refinement techniques referred to above also work best in this kind of coordinate system.

To solve the boundary condition problem, one may need a radical reformulation of the basic equations. One approach, devised originally by H Friedrich,\cite{Friedrich} is being investigated intensively by a group of scientists at the AEI and their collaborators at other institutions.\cite{Husa} In this idea, the field equations are expressed in a conformally compactified way, which means that the spacetime is mapped onto a computational domain in which the outgoing light rays reach ``infinity'' in a finite coordinate distance. There is a natural boundary condition there, and the result is a simulation of the whole spacetime with automatically varying refinement. As promising as this is in the long run, it has only recently begun and will require a number of years to reach the level of robustness of the standard spacetime-slicing approaches.

Instabilities still plague the current generation of simulations, although they grow much more slowly than in earlier codes (say,  5 years ago), thus permitting longer runs. Some seem to be associated with violations of the constraint equations, which must hold for any valid solution. Thus, methods are being studied that re-solve the constraints every time-step, which is considerable work but may be worthwhile. Another possibility is that at least some instabilities comes from bad coordinate choices, so there is considerable experimentation on this front. 

\begin{figure}[b]
\begin{center}
\includegraphics[width=.6\textwidth]{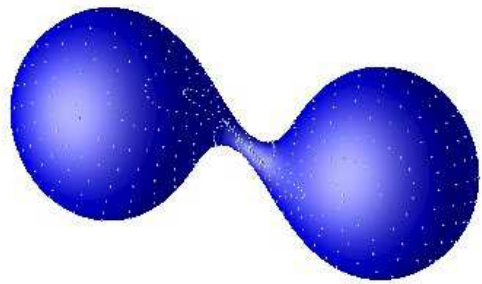}
\end{center}
\caption[]{One frame from a reconstruction of the true horizons during a black hole merger simulation. The holes are orbiting as they merge. Courtesy P Diener.}
\label{fig:horizon}
\end{figure}

Visualization is constantly being improved. A new generation of horizon finders has joined the Cactus code.\cite{Thornburg} Finding the black hole horizon is not a trivial task, because the horizon is not locally defined. It consists of the light-rays that are marginally trapped after the entire evolution has settled down. Thus, one finds the horizon by ``shooting'' photons backwards in time to find the last one that manages to escape. This requires one to save the geometry at all the relevant time-steps, for example, which adds to the computational problems. Figure~\ref{fig:horizon} below is a visualization of the result of this horizon finder.

Finally, recognizing that collaboration is ever more important in this field, numerical relativists have been among the leaders in the Grid computing movement. The GridLab collaboration,\cite{GridLab} supported by an EU research grant, has developed a suite of tools that allow Cactus users to access computing resources all over the world. Another EU grant supports a network of scientists working together on gravitational radiation source problems.\cite{GWNetwork} In Germany, a special research grant links the AEI with the Universities of Hannover, Jena, and T\"ubingen in studies of gravitational radiation, including numerical studies.\cite{SFB} 

\begin{figure}[b]
\begin{center}
\includegraphics[width=.8\textwidth]{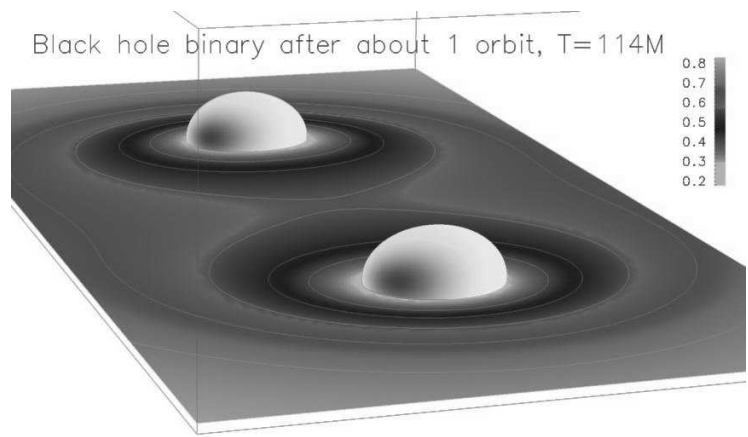}
\end{center}
\caption[]{The state of the Br\"ugmann et al \protect\cite{Bruegmann} simulation after one complete orbit. Courtesy B Br\"ugmann.}
\label{fig:bruegmann}
\end{figure}

\section{Key issue: the problem of initial data}
A numerical simulation must start with a representation of the black holes at some point late in their inspiral phase. Since our knowledge of their location at this time is a result of solving the post-Newtonian approximation, we do not have a complete description of the spacetime metric at this initial time. In particular, the initial gravitational-wave content of the metric is poorly known. This is made worse by the fact that our current methods for solving the constraint equations at the initial moment of time make certain simplifying assumptions that essentially freeze the gravitational wave content in an uncontrollable way. There is thus the possibility that the initial configuration for the numerical integration does not represent two black holes after a long inspiral phase. 

One way around this would be to let the holes orbit once numerically, during which any unwanted gravitational radiation (and also any other unwanted irregularities imposed by the initial value formulation, such as distorted horizons) would have time to go away. At present we do not have the luxury of such long integration times, but in the future this may well be the way to cure this aspect of the initial-value problem.

However, a far more worrying aspect of the problem faces groups today. Simply put, they don't know where to start the holes and what velocities to give them. There is no analytic solution to the Kepler problem, so when a group decides to start with holes at a separation of, say, $7M$, they have no exact guide to what circular velocities to give the holes to ensure that they are actually in a quasi-circular orbit. Moreover, they do not even know at present whether $7M$ is inside or outside the ISCO. If it is inside, then the holes will plunge together rapidly. If it is outside, then they may orbit stably until the group runs out of computer time! And we don't yet know whether the ISCO is so well-defined: maybe there is a broad region where the circular orbit begins to go unstable. Even if we knew where the place the holes to start them off, the uncertainty in their orbital speeds would also be a problem. Taking a speed smaller than they would actually arrive with would again cause them to plunge together too rapidly. Taking too large a speed would cause them to separate again. 

While it is possible to explore these issues numerically, this would take a huge amount of computer resource. At present several groups are trying to use heuristic models of the circular-orbit problem as a guide to the right data.\cite{Buonanno} And numerical data found by a new technique based on spectral methods seems to be in good agreement with these heuristics.\cite{Meudon2} But this problem is far from being solved, and until we have a better understanding of it, it will be difficult to trust any waveform predictions.

\section{Numerical evolutions}
The most ambitious black hole simulation attempted to date was the AEI's ``Discovery Channel'' simulation, so-called because it was performed and visualized for a program aired by the Discovery Channel network in 2003. It followed two equal-mass Schwarzschild black holes from what was then our best guess as to the location and orbital velocities of the ISCO. The simulation showed that the holes immediately plunged together and merged in less than half an orbit. This was not expected, and it suggests that the initial position was actually inside the ISCO and/or that the initial velocities were too small. 

The evolution itself was very successful, and the group could follow the merger through ringdown of the final black hole. The excision regions remained stable throughout. The merger of the horizons was also very smooth, as has been shown by Diener: see Figure~\ref{fig:horizon}.

The record for the longest stable simulation of two black holes in orbit belongs was set late last year with a simulation by Br\"ugmann and collaborators.\cite{Bruegmann} Figure~\ref{fig:bruegmann} shows the horizons of the black holes {\em after} one complete orbit. They show little distortion and have kept their distance well. The simulation used the group's own mesh refinement code and a number of advances devised jointly with the AEI group. The orbit took $114M$ of time, and the simulation actually remained stable for $150M$. Even more remarkably, it took only 24 hours to perform on a single processor! The performance of a single stable circular orbit is indeed a major step forward in the current state of this field!

More groups can be expected to attempt to emulate the success of this simulation, both using finite differences and in the realm of spectral methods. It encourages me to be optimistic that in two to three years we will have reasonably accurate waveforms for extracting the merger signals of non-spinning black holes from the data of ground-based gravitational wave detectors. Hopefully, that will be rapidly followed by further improvements. Spin will be important for the merger signal, and most merging black holes can be expected to have a substantial spin. And even more important, especially for LISA, will be simulations of unequal-mass mergers. These are currently being developed in a number of groups. 

Nevertheless, progress in this field will not speed up overnight. It is limited by a shortage of people, and in recognition of this, NASA and NSF have been discussing a joint initiative to fund more effort in the USA. If this should materialize, it would bring the day when these simulations provide really useful results considerably closer.

%

\end{document}